# Ion Transport Mechanisms in Pectin-containing EC-LiTFSI Electrolytes

*Sipra Mohapatra*[1], *Hema Teherpuria*[1], *Sapta Sindhu Paul Chowdhury*[1], *Suleman Jalilahmad Ansari*[1], *Prabhat K Jaiswal*[1], *Roland R. Netz*[2], *and Santosh Mogurampelly*[1,2*]

[1]Department of Physics, Indian Institute of Technology Jodhpur, Karwar, Rajasthan 342037, India.

[2]Fachbereich Physik, Freie Universität Berlin, 14195 Berlin, Germany.

*Corresponding Author: santosh@iitj.ac.in

## Abstract

Using all-atom molecular dynamics simulations, we report the structure and ion transport characteristics of a new class of solid polymer electrolytes that contain biodegradable and mechanically stable biopolymer pectin. We simulate a highly conducting ethylene carbonate (EC) as a solvent for lithium-trifluoromethanesulfonimide (LiTFSI) salt containing different weight percentages of pectin. Our simulations reveal that the pectin chains reduce the coordination numbers of lithium ions around the counterions (and vice-versa) because of stronger lithium-pectin interactions compared to lithium-TFSI interactions. Further, the pectin is found to promote smaller ionic aggregates over larger ones, contrary to results typically reported for liquid and polymer electrolytes. We observe that the loading of pectin in EC-LiTFSI electrolytes increases the viscosity ($\eta$) and relaxation timescales ($\tau_C$), indicating higher mechanical stability and, consequently, the mean squared displacements, diffusion coefficients ($D$), and the Nernst-Einstein conductivity ($\sigma_{NE}$) decrease. Interestingly, while the lithium diffusivities are related to the ion-pair relaxation timescales as $D_+ \sim \tau_c^{-3.1}$, the TFSI$^-$ diffusivities exhibit excellent correlations with ion-pair relaxation timescales as $D_- \sim \tau_c^{-0.95}$. On the other hand, the NE conductivities are dictated by distinct transport mechanisms and scales with ion-pair relaxation timescale as $\sigma_{NE} \sim \tau_c^{-1.85}$.

## Introduction

The nature of electrolyte material sandwiched between the electrodes in a rechargeable battery device is highly critical for developing advanced lithium-ion batteries that are safe, lightweight, durable, and mechanically stable.[1–4] Among different classes of electrolyte materials, solid polymer electrolytes (SPEs) are generally preferred due to several advantages.[2,4,5] Explicitly, the SPEs outperform the traditionally used liquid electrolytes in providing safety, better solid electrolyte interface, and better mechanical stability to the batteries stopping dendrite growth.[6–9] The SPEs also serve as a better choice over liquid electrolytes due to their high mechanical stability, ease of processing, and negligible interfacial resistance.[1,10] Accordingly, significant efforts are focused on addressing some of the critical issues associated with SPEs, including their low ionic conductivity at room temperature in fully commercializing the SPE-based battery technologies.

Besides increasing ionic conductivity, developing advanced electrolyte materials which are also biocompatible holds promise for future battery technologies. Due to their solid nature, biocompatible solid electrolytes have the significant advantage of being chemically stable and have minor safety issues.[11,12] In this context, researchers are actively exploring advanced SPEs based on naturally occurring biopolymers instead of synthetic polymers (such as PEO) due to the promise of environmental compatibility owing to their biodegradable nature.[9,13–15] Naturally occurring polymers are mainly composed of repeated monomeric units of saccharides, fatty acids, amino acids, and nucleotides and are primarily found in biological organisms.[16,17] Polysaccharides are natural biopolymers made of carbohydrate monosaccharides linked through o-glycosidic linkages. Biopolymers such as starch, chitosan, agar-agar, cellulose, etc., are previously proposed as polymer hosts in battery electrolytes.[8,14,18,19] In the past decade, significant progress has been made in improving the properties of biocompatible electrolyte materials for battery applications. For instance, biodegradable gel polymer electrolytes developed by Gou et al. showed stable capacity retention, minimal fading, and high Coulombic efficiency.[20] Lin et al. designed biocompatible electrolytes using corn starch within the PEO-LiTFSI system in different wt. %s of corn starch.[21] They observed an increase in ionic conductivity of the composite electrolyte with the loading of corn starch and the highest conductivity of $2.62\times10^{-5}$ S/cm at a ratio of PEO:corn starch=9:1. Biocompatible SPEs developed by Zhou et al. are stretchable and flexible with a self-healing nature that can be crucial in battery applications.[22] For the commercial adoption of biocompatible SPEs in batteries, the ionic conductivity should be comparable to that of liquid electrolytes. However, the ionic conductivity of biocompatible electrolytes is typically much lower than that of liquid electrolytes ($10^{-2}$–$10^{-3}$ S/cm)[23,24], which remains a significant challenge for their applications. Since the ionic conductivity of the polymer-based electrolytes is highly dependent on the ion-solvating capabilities of the host polymer matrix, the choice of biopolymer is critical for developing efficient battery electrolyte materials.

Pectin ($C_6H_{10}O_7$), a polysaccharide biomacromolecule consisting of α−1,4 linked D-galacturonic acid monomers, is highly abundant in nature and typically found in the cell walls of plants, fruit extracts, and agricultural products.[25,26] Pectin plays a key role in maintaining the cell wall structure and provides rigidity to plants and fruits. Further, pectin indirectly helps in regulating the permeability of water, nutrients, and other small molecules in the biological environment. Some environmental advantages of using pectin in technological applications include the polysaccharide nature, ease of fabrication, and ultralow toxicity.[19] Considering these intriguing features, we propose to integrate pectin with traditional highly conducting liquid electrolytes for the development of advanced solid-state battery electrolyte technologies. Besides possessing the desired solvating capabilities and environmental advantages, different functional groups on monomers can result in a variety of structural complexities of pectin that are likely to offer exciting consequences for ion transport. One of the possible outcomes of the flexibility in choosing different monomeric architectures is the pectin-based SPE with unique ion transport characteristics similar to those reported for single-ion conductors and/or polymeric ionic liquid electrolytes.[27,28] The anionic nature of the monomeric polysaccharide unit enables ionic coordination with the surrounding cations present in the electrolyte material, similar to polyelectrolytes,[29–31] and may promote efficient ion transport.[24] Motivated by the above-discussed issues, we propose pectin biomacromolecules for developing pectin-based SPEs as alternatives to synthetic polymer-based SPEs and investigate their structure and ion transport properties.

There are only a few experimental studies about pectin-based electrolytes in the literature. In 2009, J. R. Andrade et al. prepared a transparent gel electrolyte composed of methanol-esterified pectin chains and glycerol as a plasticizer with $LiClO_4$ salt.[32] They observed the ionic conductivity of methanol-esterified pectin-based glycerol-$LiClO_4$ gel bio-polymer electrolyte to be $4.7\times10^{-4}$ S/cm at 68 wt. % of glycerol at room temperature. Perumal et al., in their experimental work,[33] found that the electrolyte composed of equal molecular wt.% of pectin and LiCl offered ionic conductivity of $1.96\times10^{-3}$ S/cm, which is higher than the $LiClO_4$ (40 wt. % $LiClO_4$+60 wt. % pectin) incorporated electrolyte with ionic conductivity of $5.38\times10^{-5}$ S/cm. A thorough understanding of the structural properties, underlying transport mechanisms, and mechanical stability of pectin-based SPEs are still required to have an alternative to the traditionally used electrolytes.

In this work, we investigate the effect of pectin loading on the transport and structural properties of typical commercial EC-LiTFSI battery electrolytes using atomistic simulations. The ion structure is investigated to study the ion-solvating capability of pectin, which is a crucial factor for the transportation of ions in electrolytes. The radial distribution function (g(r)), coordination number (CN(r)), and ion association probability (P(n)) are investigated to study the structural compactness and intermolecular interaction between the ion-ion, ion-polymer, and ion-solvent. The underlying ion transport mechanism is quantified through the ion-pair relaxation time ($\tau_c$) and the diffusion coefficient ($D$) of the Li cation and TFSI anion. Further, we calculate ionic conductivity and cationic transference to understand the ionic transport behavior of the pectin-loaded systems. Finally, we summarize our results and in relation to the ion transport and the mechanical stability of EC-loaded with different wt. % of pectin.

## Simulation Methods

### Interaction Potential Model and Force Fields for Pectin and EC-LiTFSI Electrolytes

We perform all-atom molecular dynamics simulations of pectin-EC-LiTFSI electrolytes at different loading of pectin using GROMACS 2021.2 package[34,35] with the following interaction potential:

$$U(r) = U^{\text{bonded}}(r) + \sum 4\epsilon_0 \left[ \left( \frac{\sigma}{r_{ij}} \right)^{12} - \left( \frac{\sigma}{r_{ij}} \right)^{6} \right] + \sum \frac{q_1 q_2}{4\pi\epsilon_0 r_{ij}}. - - - - - - (1)$$

In the above equation, the $U^{\text{bonded}}$ includes all the intramolecular interactions with contributions from bonds, angles, and torsions. The remaining terms correspond to the nonbonded interactions and are modeled with the Lennard-Jones interaction potential and the Coulomb potential. A scaling factor of 0.5 is used for the nonbonded interactions between intramolecular atomic pairs separated by three bonds. The scaling factor, however, is not used for the intramolecular atomic pairs separated by more than three bonds. A real space cutoff of 12 Å is used for the LJ and Coulomb interactions, and k-space summation for the electrostatic interactions is carried out using the particle mesh Ewald method.[36] The LJ interactions beyond the cutoff distance were truncated by including analytical tail corrections for pressure and energy.

The force field parameters for EC and LiTFSI salt are excerpted from the standard optimized potential for liquid simulations–all-atom (OPLS-AA) force field set developed by Jorgensen[37] with improved intramolecular parameters from Acevedo.[38] The total charge on ionic species was scaled to 0.8e to indirectly mimic the induced polarization effects in a mean-field-like manner.[39] This approach was previously shown to produce results compared to polarizable models and experiments.[40–44] The OPLS-AA parameter set produces good results for the structure and diffusion coefficient of ions in neat EC-LiTFSI electrolytes consistent with experiments.[5,45] The LJ nonbonded parameters for all atomic types in pectin are taken from the GLYCAM06J parameter set[46] suitable for polysaccharides. Among different esterification states of pectin [depending on the extraction from agricultural products and possessing different properties], we considered the COOH esterification on its monomers. Therefore, the intramolecular interaction parameters (excluding the dihedral angles) and partial atomic charges of pectin are developed in this paper by performing quantum mechanical calculations, as explained in the following **Section**. The Lorentz-Berthelot arithmetic rules were followed to calculate the nonbonded parameters for the cross-interaction terms between different atom types of pectin. However, to be consistent with the OPLS-AA force field, the cross-interaction terms between different atomic types of EC-LiTFSI were calculated using the geometric mixing rules. Similarly, cross-terms between pectin and EC-LiTFSI were also calculated using the geometric mixing rules.

**Development of the Intramolecular Force Field Parameters for Pectin**

Quantum mechanical calculations were performed to generate the forcefield parameters by simulating the monomer of pectin using the density functional theory (DFT) with B3LYP/6-311g** basis set[47,48] using the Gaussian16 package.[49] We have optimized the geometry of a monomeric unit of pectin to calculate the equilibrium bond lengths and angles in pectin.[50] The normal mode analysis of equilibrium vibrational frequencies was performed to estimate the force constants of the harmonic potentials for all bonds and angles. The partial atomic charge of all atomic types of pectin was evaluated by fitting the electrostatic potential to atomic centers using the RESP fitting method.[51] The parameters of equilibrium dihedral torsion angles and the respective force constants are taken from the GLYCAM06J parameter set.[46] The complete force field parameters are provided in **Table ST1** in the supplementary information (SI).

**Initial System Setup**

The molecular structure of the pectin chain was generated using GLYCAM oligobuilder software, consisting of 12 monomeric units.[46] The chemical composition of the monomeric unit is 1−4α linkage of D-galacturonic acid ($OH-[C_6H_{10}O_7]_n-OH$),[26] and the polymer chain was terminated with a hydroxyl group. Pectin biopolymer is a class of wall polysaccharides with a complex structure. According to their functionality, various pectin structures are available, like homogalacturonans and rhamnogalacturonans acidic polymers. Esterification is attaching an ester (R−OOH) group to the pectin chain. The homogalacturonans are the linear combination of D-galacturonic acid with 1-4 alpha linkage, which we have used to investigate the movement of ions in a liquid-based system.[52]

The initial configuration of the electrolyte system was then prepared by randomly placing 5240 EC molecules in the simulation box using PACKMOL.[53] The EC system was then solvated with 100 $Li^+$ and 100 $TFSI^-$ ions to maintain an approximately constant salt concentration for the LiTFSI salt in all the electrolytes studied in this paper. To prepare different composite electrolyte systems with varying wt. %s of pectin, we added the appropriate numbers of pectin chains in the simulation box. We prepared 8 different composite electrolyte systems with pectin wt. %s of 0.5, 1, 2, 3, 5, 10, 30, and 50, along with the neat EC-LiTFSI (0 wt. %). To prepare the initial system with a tolerance of 2 Å between any two atoms. Further, such a low density was chosen to ensure that the electrolyte systems do not suffer potential energy traps leading to numerical instabilities arising from the close contacts between different atoms or molecules. Details of the number of atoms, volume, density, etc., of different pectin wt. %s in EC-LiTFSI systems at 425 K are reported in **Table ST2** in SI.

**Equilibration Protocol**

Minimization was performed on the systems consisting of pectin with different wt. %s solvated with suitable numbers of $Li^+$ cations and $TFSI^-$ anions and EC molecules. The PACKMOL[53] generated structures containing pectin, EC, and LiTFSI were minimized using the steepest descent method for 1000 steps with a tolerance of 10 kJ/mol/nm on forces. The energy minimized systems were then subjected to an NVT ensemble using a V-rescale[54] thermostat for 100 ps with 1 fs step size and subsequently subjected to a 10 ns NPT ensemble with a V-rescale thermostat and Berendsen barostat[55] with a timestep of 1 fs at 425 K. The damping relaxation times of 0.1 ps and 2 ps were used in the thermostat and barostat, respectively. A compressibility factor $4.5\times10^{-5}$ $bar^{-1}$ was used for the NPT ensemble. Bond lengths involving hydrogen atoms were constrained to their equilibrium lengths using the LINCS algorithm.[56] The equations of motion are integrated using the leapfrog algorithm with a simulation timestep of 2 fs,[57] and the trajectories were saved every 1 ps. A representative snapshot of pectin-EC-LiTFSI electrolytes at 30 wt. % was shown in **Figure S1** to give a visual impression.

As discussed above, the equilibration protocol consists of initial minimization, then a short NVT, and finally, the density equilibration with an NPT ensemble. After attaining the equilibrium, the simulated density of the neat electrolyte (EC-LiTFSI, EC:Li=48:1, 425 K) was obtained to be $1186 \pm 5\ \mathrm{kg/m^3}$, comparable to experiments.[58–60] We generated 300 ns long production trajectories at different wt. %s of pectin in the NPT ensemble, among which the last 50 ns was used for analyzing the structural properties (such as g(r), CN(r), and P(n)), and the entire trajectory was used for analyzing the transport and relaxation phenomena (such as MSDs, $D$, $\tau_C$, and $\sigma_{NE}$). For viscosity calculations, we performed 50 independent NPT runs of 1 ns with a finer resolution saving frequency of 1 fs to calculate the pressure tensor autocorrelation function.

## Results and Discussion

Here, we discuss the structural and transport properties of pectin and ions in pectin-EC-LiTFSI electrolytes. The ion diffusion coefficients are compared with ion-pair relaxation timescales to understand the underlying ion transport mechanisms.

### Ion Density Profiles

As discussed in the introduction, we propose/hypothesize pectin to possess good ion-solvating capabilities for its applications in lithium-ion batteries. To examine the nature of ion-solvating features, we calculate the density profile of different ionic species in pectin-loaded EC-LiTFSI electrolytes in the x, y, and z directions. As shown in **Figure 1**, for the neat EC-LiTFSI system (0 wt. %, bottom subplot), we confirm the uniform distribution of $Li^+$ and $TFSI^-$ ions in the electrolyte. This result establishes that EC is an excellent solvent for uniformly dissolving LiTFSI salt. With the addition of pectin to EC-LiTFSI electrolytes, the uniform distribution of ions is not significantly affected, but the distribution appears to fluctuate, which increases with the loading. These fluctuations with higher loading are expected because of the electronegative charge groups on the backbone of pectin. The above results suggest that no ionic clusters are formed in the presence of pectin, supporting ion conduction and the suitability of pectin for electrolyte applications.

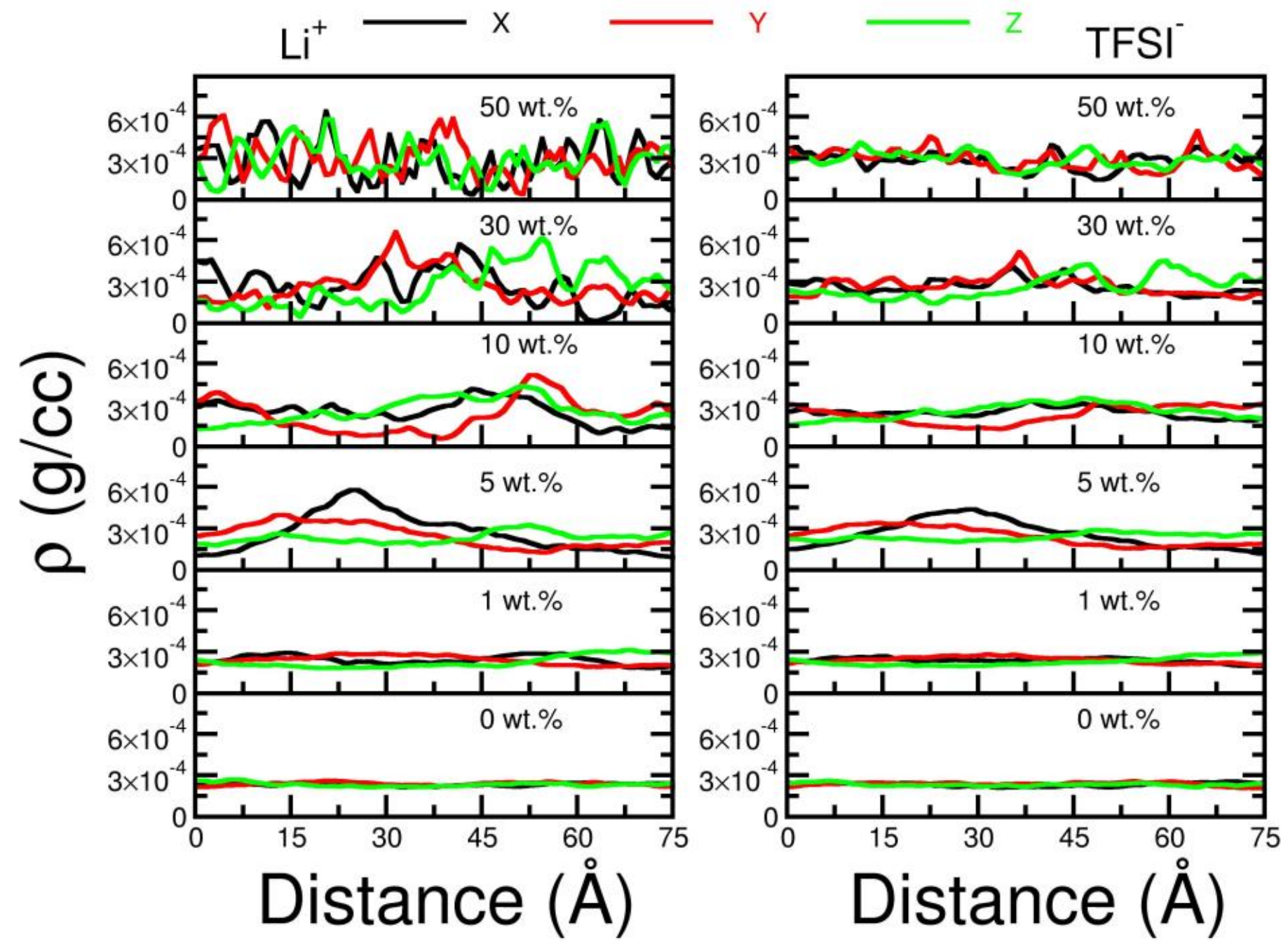

**Figure 1**: Number density profile of $Li^{+}$ and $TFSI^{-}$ ions in pectin-loaded systems at different wt.% along the x, y, and z axes.

**Radial Distribution Functions and Association Probability**

Apart from the ion-solvating capabilities, the uniform distribution of ionic species and associated fluctuations also indicates the abundance of strong ion-ion and ion-polymer interactions.[61] To understand the structural arrangement of the pectin-EC electrolytes, we calculated the radial distribution functions, $g(r)$, between different atomic species of ion-ion and ion-polymer interactions and the respective coordination numbers, $CN(r)$. **Figure 2(a)** and **2(b)** display the $g(r)$ and $CN(r)$ between the anionic and cationic pairs. We observed the first peak and first minima at distances of 4.25 Å and 5 Å, respectively. However, a monotonic decrease in the first peak is observed with the loading of pectin chains. These results indicate that the ion-pair coordination in neat EC-LiTFSI electrolytes is relatively strong despite their high ionic conducting properties. Further, the monotonic decrease of ion-pair correlations with the pectin loading suggests that pectin can facilitate efficient ionic conductivity compared to electrolyte systems dominated by strong ion-pair correlations.[62] Despite the strong nature of ion-ion interactions, the $CN(r)$ reveals that roughly half an anion (~0.6 numbers of $TFSI^{-}$ ions) is coordinated around cations within the first coordination shell in neat EC-LiTFSI electrolytes, consistent with previous calculations, which further reduced with the loading of pectin.

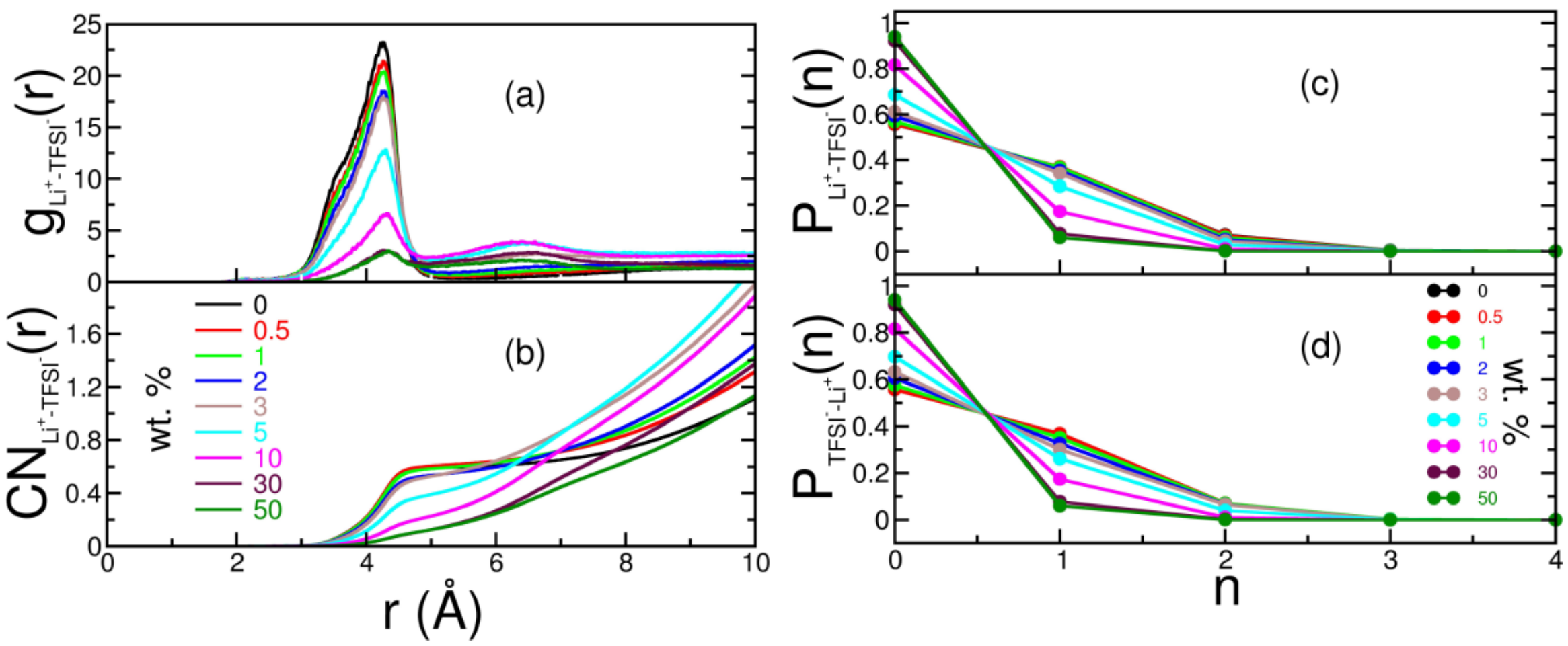

**Figure 2**: (a) Radial distribution function of Li-TFSI ionic pairs, (b) coordination number of $TFSI^-$ around $Li^+$ ions, and ion association probability of (c) finding $n$ number of $TFSI^-$ ions around $Li^+$ ions, and (d) finding $n$ number of $Li^+$ around $TFSI^-$ ions.

For a detailed understanding of the probability of either free ion formation or n number of ion coordination around their counterions, we calculate the ion association statistics, ($P(n)$) and show the results in **Figure 2(c)** and **2(d)**. The ion association probability is calculated using,

$$P(n) = \frac{1}{N_{\text{frames}}} \sum_{i=1}^{N_{\text{frames}}} \sum_{j=1}^{N_{\text{ions}}} \frac{\delta_{n_i n_j}}{N_{\text{ions}}}, ------ (2)$$

where $\delta_{n_i n_j}$ is the Kronecker delta function to count $n$ number of counterions within the first coordination shell of the ion $j$, $N_{\text{ions}}$ is the total number of ions, $N_{\text{frames}}$ is the total number of frames. Here, $\delta_{n_i n_j}$ is defined in such a way that $\delta_{n_i n_j} = 1$ if $n_i^{\text{th}}$ ion is found within the first coordination shell of $n_j^{\text{th}}$ ion, and $\delta_{n_i n_j} = 0$, otherwise. Consistent with $g(r)$ and $CN$(r), the probability of finding free ions increase with pectin loading. Consequently, the probability of $n$=1 or 2 ions around their counterions decreases with pectin. This result may be explained as a consequence of the dilution effect due to the loading of pectin (i.e., loading of pectin to EC-LiTFSI decreases salt concentration). Despite the dilution due to pectin, the decrease in small ionic aggregates at high loading of pectin is a very promising result for efficient ionic conductivity.[63] Therefore, the presence of pectin chains helps in decoupling the ion-ion correlations that would promote efficient ionic conductivity.

The $g(r)$ and $CN(r)$ associated with different atomic pairs representing the ion-polymer and polymer-polymer interactions are provided in the SI (**Figure S2**). The ion association probabilities of other important atomic species are shown in **Figure S3** in the SI. Briefly, these results indicate strong polymer-polymer interactions and polymer-cation interactions in the pectin-EC electrolytes. The implications of these results on ion transport and viscosity are discussed in the following sections.

**Mean Squared Displacement and Diffusion Coefficient**

The present work is based on the hypothesis that the involvement of the pectin matrix can affect the ion conductivity and mechanical stability of SPEs. To clarify this, we have calculated the mean squared displacement to study the ion transport properties of both cations and anions. The self-diffusion coefficient can help us understand ion transport characteristics, and the diffusion coefficient can be determined from the mean squared displacement (MSD). We have calculated the MSD for both $Li^+$ and $TFSI^-$ ions from,

$$\mathrm{MSD}(t) = \frac{1}{N_{\mathrm{frames}} - t} \sum_{t'=0}^{N_{\mathrm{frames}}-t} \frac{1}{N} \sum_{i=1}^{N} [\vec{\boldsymbol{r}}_i(t+t') - \vec{\boldsymbol{r}}_i(t')]^2, ------ (3)$$

where $N$ is the number of atoms and $\vec{\boldsymbol{r}}_i$ is the position vector. The results shown in **Figure 3(a)** indicate that lithium ions remain in the subdiffusive region up to an order of $10^4$ ps. A clear diffusive regime is observed at timescales beyond 100 ns for lower wt. %. The extent of the subdiffusive regime and the onset of the diffusive regime increase with the loading. Interestingly, the $TFSI^-$ ions show higher mean squared displacements than lithium ions, with the distinction being more pronounced at higher loading. Since the $TFSI^-$ ions are large molecular ionic species than lithium ions, we propose that a loosely packed solvation shell caused due to the delocalized ionic charge distribution around $TFSI^-$ ions help them display higher MSDs in relation to lithium ions, inspired by recent works of Olvera de la Cruz[64] and Hall.[65] However, while we do not have clear evidence for the correlation between ion diffusion and its size, there is still debate on whether the size of an ion has a direct connection with the ionic diffusivity.[66,67] As seen in **Figure 3(a)**, we observe MSDs to decrease with pectin loading with implications for the diffusion coefficient of ions, which is calculated as $D = \lim_{t\to\infty} D^{\mathrm{app}}(t) = \lim_{t\to\infty} \frac{\mathrm{MSD}(t)}{6t}$, where $D^{\mathrm{app}}(t)$ is the time dependent apparent diffusion coefficient calculated from the diffusive regime of MSD curves (see **Figure 3(b)**). In **Table ST3** of the SI, we provide the details of the time period over which the MSD curves were fitted to calculate the diffusion coefficient of ions and the diffusion exponent as in $\mathrm{MSD}(t) \sim t^{\lambda}$ obtained from the apparent linear regime.

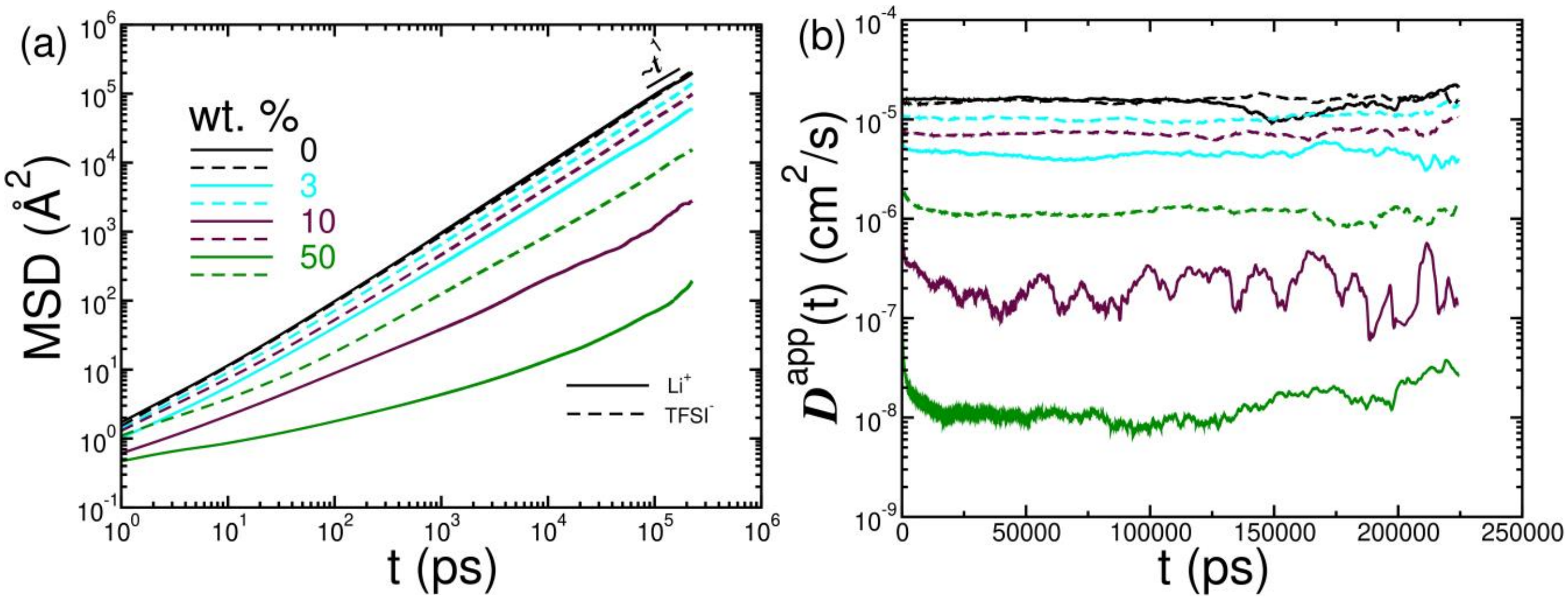

**Figure 3**: (a) MSD of $Li^+$ and $TFSI^-$ at selected values of pectin loading, i.e., 0, 3, 10, 50 wt. %s loading of pectin chains. The $TFSI^-$ ions show higher MSD than $Li^+$ ions, and (b) the apparent diffusion coefficient as a function of lagtime at different wt. % of pectin loading. (legends are same as **Figure 3(a)**)

In **Figure 4**, we display the diffusion coefficient of the cations and anions as a function of loadings of the pectin chains. Because pectin is a highly viscous material, the diffusion coefficient of ions is expected to decrease, and the effect increases with the increasing loading of pectin. Consistently, we find that the diffusion coefficient of ions decreases in the presence of pectin. The diffusion coefficient of $TFSI^-$ ions is higher than that of $Li^+$ ions, consistent with the corresponding MSD data presented in **Figures 3(a)** and **S4**. However, we observe an interesting trend at lower pectin content (up to 3 wt. %) that the rate of change of diffusion coefficient with pectin loading is somewhat similar for both the $Li^+$ and $TFSI^-$ ions. Further, the diffusivity of lithium ions is more significantly affected compared to the $TFSI^-$ ions. The decay rate in diffusivity of the $Li^+$ ion is substantially faster than that of the TFSI ion, indicating a decrease in total ionic conductivity. As a monatomic ion, lithium has more tendency to interact with polymeric electronegative groups and be trapped, which explains why lithium diffusion is drastically reduced compared to the diffusion of the larger TFSI ion. Explicitly, the ratio between the diffusion coefficients of lithium and $TFSI^-$ ions is almost 100 at higher wt. %s of pectin in the electrolytes.

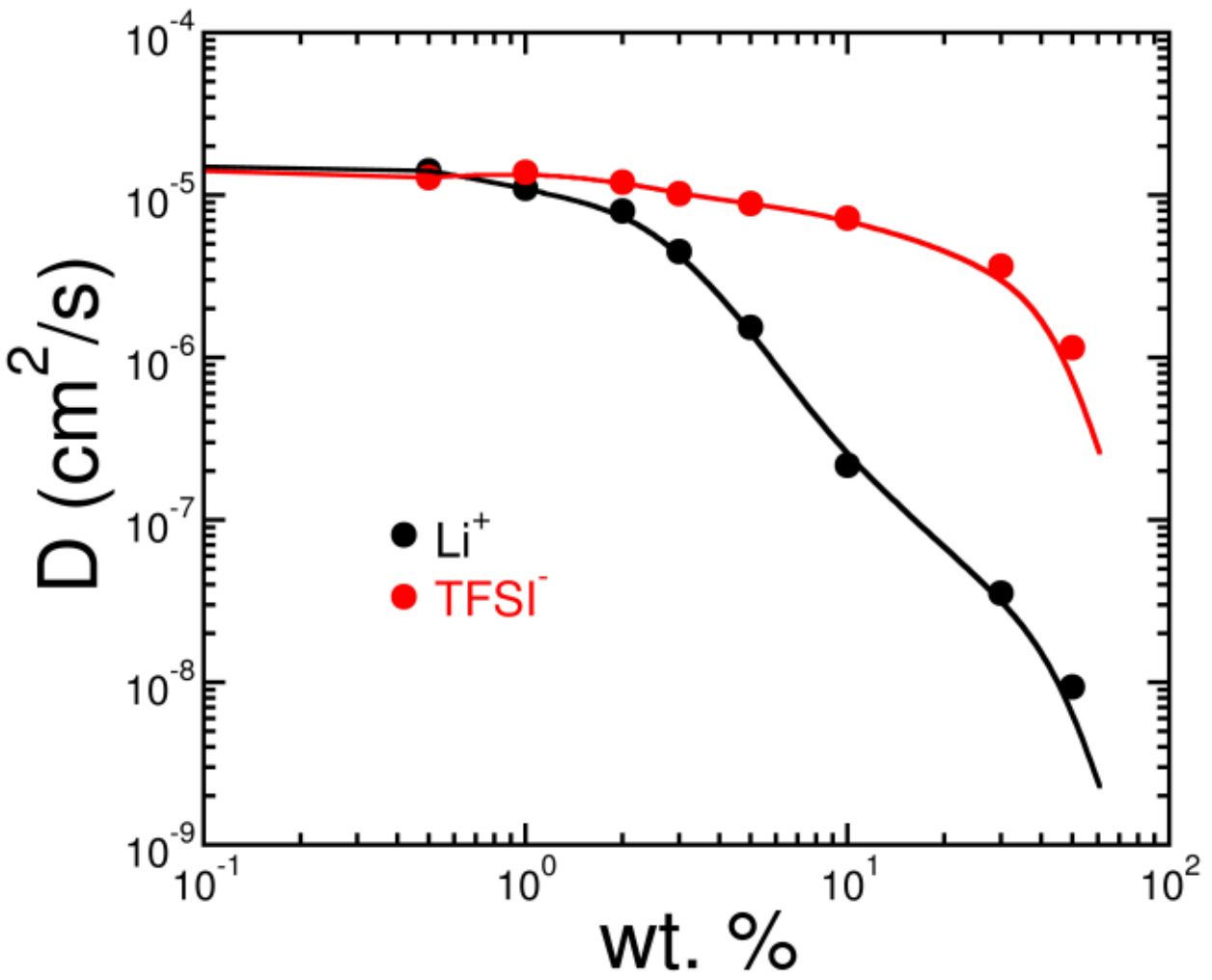

**Figure 4**: Diffusion coefficient of $Li^+$ and $TFSI^-$ ions as a function of wt. % of pectin in EC-LiTFSI electrolytes. The solid lines are spline fits meant to guide to the eye.

### Viscosity and Ion-pair Relaxation Timescales

It is known that the viscous nature of the electrolyte matrix and the underlying relaxation phenomena significantly influence ion transport in SPEs.[68–70] Specifically, in SPEs containing polyethylene oxide, it was well established that diffusivity correlates excellently with viscosity and polymer dynamics.[43,71–73] Similarly, viscosity and the ion-pair relaxation phenomena typically dictate the diffusivities in most liquid types of electrolytes. Since our pectin-loaded EC-LiTFSI electrolytes exhibit the features of both liquid and solid types of electrolytes, it will be crucial to understand the viscosity and the relaxation phenomena. Particularly, we want to ask the question: how are ionic diffusivities related to viscosity and various relaxation phenomena? In the following sections, we shed light on this question quantitatively by directly comparing the ionic diffusivities with the viscosity and the ion-pair structural relaxation timescales.

The viscosity calculations were carried out with the Green-Kubo formula[74–77] that uses the time correlation of the pressure tensor as:

$$\eta = \frac{V}{k_B T}\frac{1}{6}\sum_{\alpha\beta}\int_0^\infty d\tau \langle P_{\alpha\beta}(t)P_{\alpha\beta}(t+\tau)\rangle, ------ (4)$$

where $P_{\alpha\beta}$ are the pressure components, $V$ is the volume, $k_B$ is the Boltzmann constant and $T$ is the absolute temperature. The ensemble average was taken over all 6 off-diagonal elements, $\alpha\beta = xy, xz, yz, yx, zx,$ and $zy,$ and 50 different trajectories were generated from completely uncorrelated initial configurations. Since the pressure autocorrelation function is highly sensitive to the saving frequency, we recorded the trajectory of $P_{\alpha\beta}$ every 1 fs. The average of pressure autocorrelation functions along with the respective 50 independent runs, corresponding running integrals, detailed discussion on reducing numerical errors, extract of the Fortran code [full version of the code is available on request] are shown in the **Figure S5-S6(b)** in the SI.

The results of viscosity as a function of pectin wt. % in EC-LiTFSI electrolytes are presented in **Figure 5(a)**. For neat EC-LiTFSI electrolytes, we find the viscosity to be 0.708±0.200 mPa.s at 425 K, off by a factor of ~3 compared to the experimental data.[78,79] The lower viscosity value compared to experiments is expected because the pair-wise interaction potential model does not capture the induced polarization effects completely, even if the scaled electrostatic interactions are employed in the potential model. Further, the viscosity changes slightly with pectin for lower loadings but rapidly for higher loadings beyond 5 wt. %. At 50 wt. %, the viscosity is calculated to be 170±142 mPa.s, about 250 times higher than the neat EC-LiTFSI electrolytes. We note that our results of viscosity at high wt. % of pectin are somewhat erroneous due to high correlations in the autocorrelation function even with a trajectory of 10 ns containing pressure tensor components recorded at every 1 fs. Consequently, more statistics is required for higher wt. % to calculate the average viscosity and to understand the mechanical strength of the system. The analysis of viscosity correlations with diffusion coefficient are presented in the **Figure S7(a)** and **(b)**.

For the pectin-EC-LiTFSI system, different relaxation timescales may prevail in the system, such as polymer dynamics, hydrogen bond timescales, and ion-pair relaxation timescales. However, since our trajectories are only 300 ns long, estimating the relaxation phenomena associated with pectin polymer dynamics is not feasible. Moreover, the $Li^{+}$ and $TFSI^{-}$ ions are less prone to hydrogen bond formation in the EC electrolytes. Therefore, apart from the viscosity, we chose to analyze the ion-pair structural relaxation phenomena and examine their connection to the self-diffusion coefficients of ions.

To quantify the ion association relaxation phenomena, we computed the ion-pair correlation function $C(t)$, which signifies the relaxation behavior of all ion associations (see **Figure S8**). The ion-pair correlation function $C(t)$ is defined as $C(t) = \langle h(t)h(0)\rangle / \langle h(0)h(0)\rangle$, where the angular bracket $\langle \cdot \cdot \cdot \rangle$ denotes an ensemble average over all ion-pairs and all possible time origins, and $h(t)$ assigned a value unity if lithium and $TFSI^{-}$ ions are found within a specified cutoff distance, and zero otherwise. The cutoff distance defining the ion-pairs is chosen as 5 Å based on the extent of the first coordination shell of lithium and $TFSI^{-}$ ions. The ion motion correlates to the average ion association lifetimes, which quantifies the dynamics of the breaking and formation of the ion association. The ion-pair structural relaxation time is calculated using $\tau_C = \int_0^{\infty} a_0 \exp\left(-(t/t^*)^{\beta}\right) \mathrm{dt} = a_0 t^* \Gamma(1 + 1/\beta)$, where $a_0$, $t^*$, and $\beta$ are the fitting parameters, and $\Gamma$ denotes the gamma function. The ion-pair relaxation time ($\tau_c$) shows a similar qualitative trend with pectin loading (**Figure 5(b)**), similar to that of viscosity. For a pure EC-LiTFSI electrolyte system, the ion-pair structural relaxation time is calculated to be 417 ps. With pectin loading, $\tau_c$ increases and reaches a maximum of 6185 ps for 50 wt. % of pectin-loaded EC-LiTFSI electrolyte systems, an order of magnitude increase compared to the pristine EC-LiTFSI system.

Our analysis of $\eta$ and $\tau_C$ concluded that while $\eta$ increases by a factor of 250 for the highest pectin loading studied in this work, the $\tau_C$ increases only by a factor of 15. Therefore, the rate of increase of viscosity and ion-pair structural relaxation time with the pectin loading follows different trends, implying that either both or none of these quantities may explain the ionic diffusivities in the electrolytes according to the Stokes-Einstein formula, $D = k_B T / 6\pi\eta R$, where $R$ is the hydrodynamic radius of the spherical particle suspended in a model fluid. To establish more clearly the relationship between the viscosity and ion-pair structural relaxation timescales, we analyzed the correlation between viscosity and ion-pair correlation timescale through $\eta \sim \tau_c^{\lambda}$ as shown in **Figure 5(c)** and obtained the exponent, $\lambda = 2.2$, which does not show a one to one correlation between $\eta$ and $\tau_c$. The above result establishes a rapid change in viscosity with the structural relaxation time. As we do not observe a one-to-one correlation between the viscosity and ion-pair structural relaxation time, we examine the effect of both the viscosity and ion-pair structural relaxation timescales on the diffusivity of ionic species.

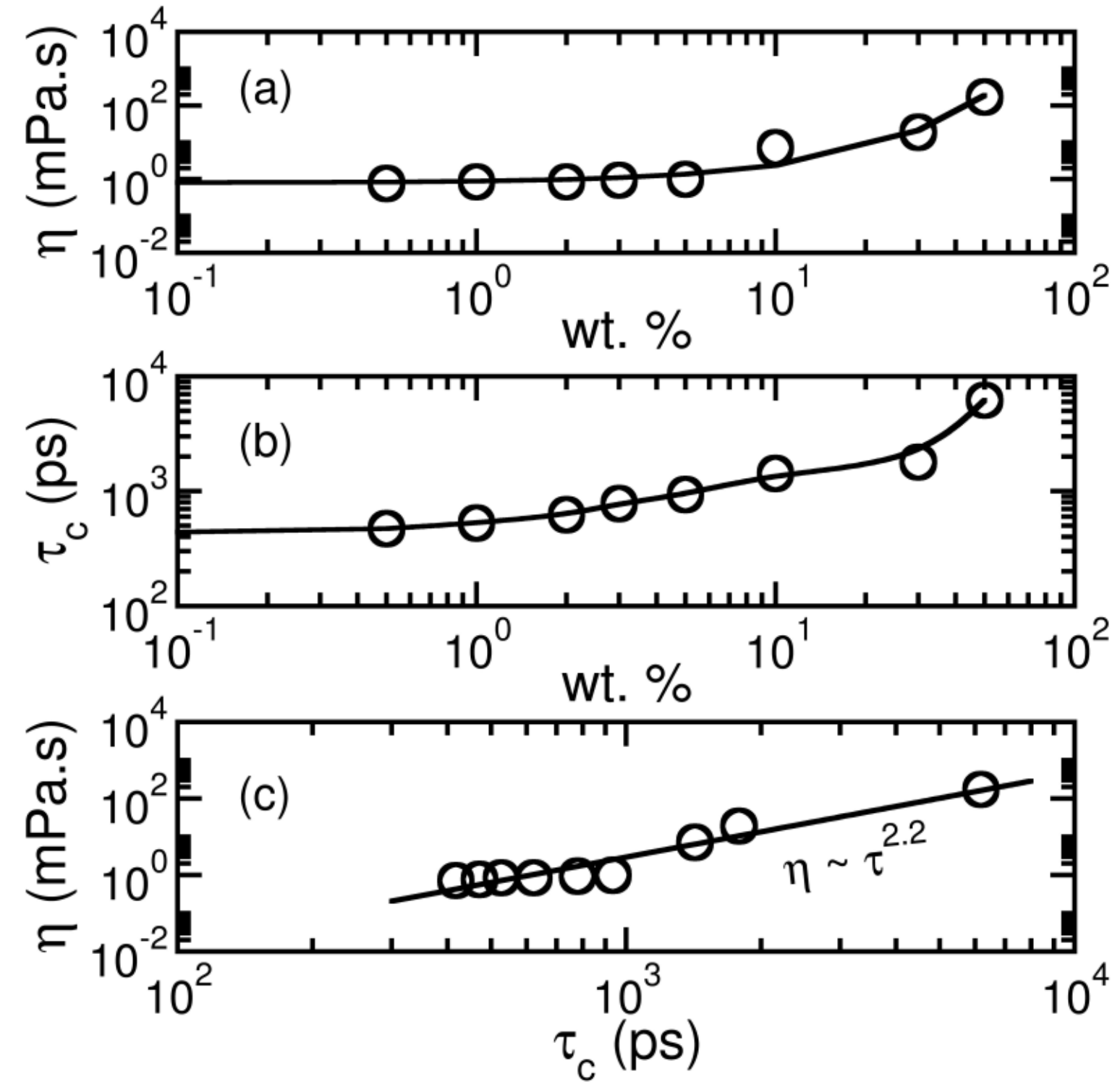

**Figure 5**: (a) Viscosity and (b) Ion-pair structural relaxation time ($\tau_C$) as a function of pectin loading. (c) Viscosity values compared against the ion-pair relaxation timescale.

In **Figures 6(a)** and **(b)**, we display the diffusivity of ions against the ion-pair relaxation timescales for $Li^+$ and $TFSI^-$ ions, respectively. For the $Li^+$ and $TFSI^-$ ions, the diffusivity decreases quite distinctly with the ion-pair relaxation timescales. Specifically, for the $TFSI^-$ ions, by fitting diffusivities to respective power laws, $D_+ \sim \tau_C^{-\lambda}$ and $D_- \sim \tau_C^{-\lambda}$, we obtain the exponents of 3.1 and 0.95, respectively. Clearly, we do not observe direct correlations between the diffusivities of $Li^+$ ions and the ion-pair relaxation timescales (i.e., $D_+ \sim \tau_C^{-3.1}$). The excellent correlations found between diffusivities of $TFSI^-$ ions and ion-pair relaxation timescales (i.e., $D_- \sim \tau_C^{-0.95}$) are similar to those reported for traditional liquid electrolytes.[44,68,70,80]

The above analysis establishes that the $Li^+$ ions are transported in the electrolyte through distinct transport mechanisms other than the underlying ion-pair structural relaxation phenomena. Since the lithium ions are strongly coordinated with the pectin chains, as noted from the respective g(r), CN(r), and P(n) plots (see **Figures S2**), we propose that pectin chains facilitate lithium ion transport along the backbone on longer timescales (likely microseconds and even longer), similar to those reported for other electrolytes containing synthetic polymers.[27,68,70,71,81] However, a further thorough analysis of long-time $Li^+$ ion transport along the pectin backbone and analyzing the respective relaxation timescales are required to establish such a hypothesis. We refrain from such an ambitious task because of the relatively short trajectories (300 ns for wt. % up to 5 and 500 ns for higher loading systems) we could generate in this work. Since $D_{+}\sim \tau_c^{-3.1}$ is also a clear indication of decoupling between the ion transport properties and the structural relaxations, there is a huge scope to achieve the twin goal of the highest ionic conductivity and mechanical stability in a single battery electrolyte.[27,68,80]

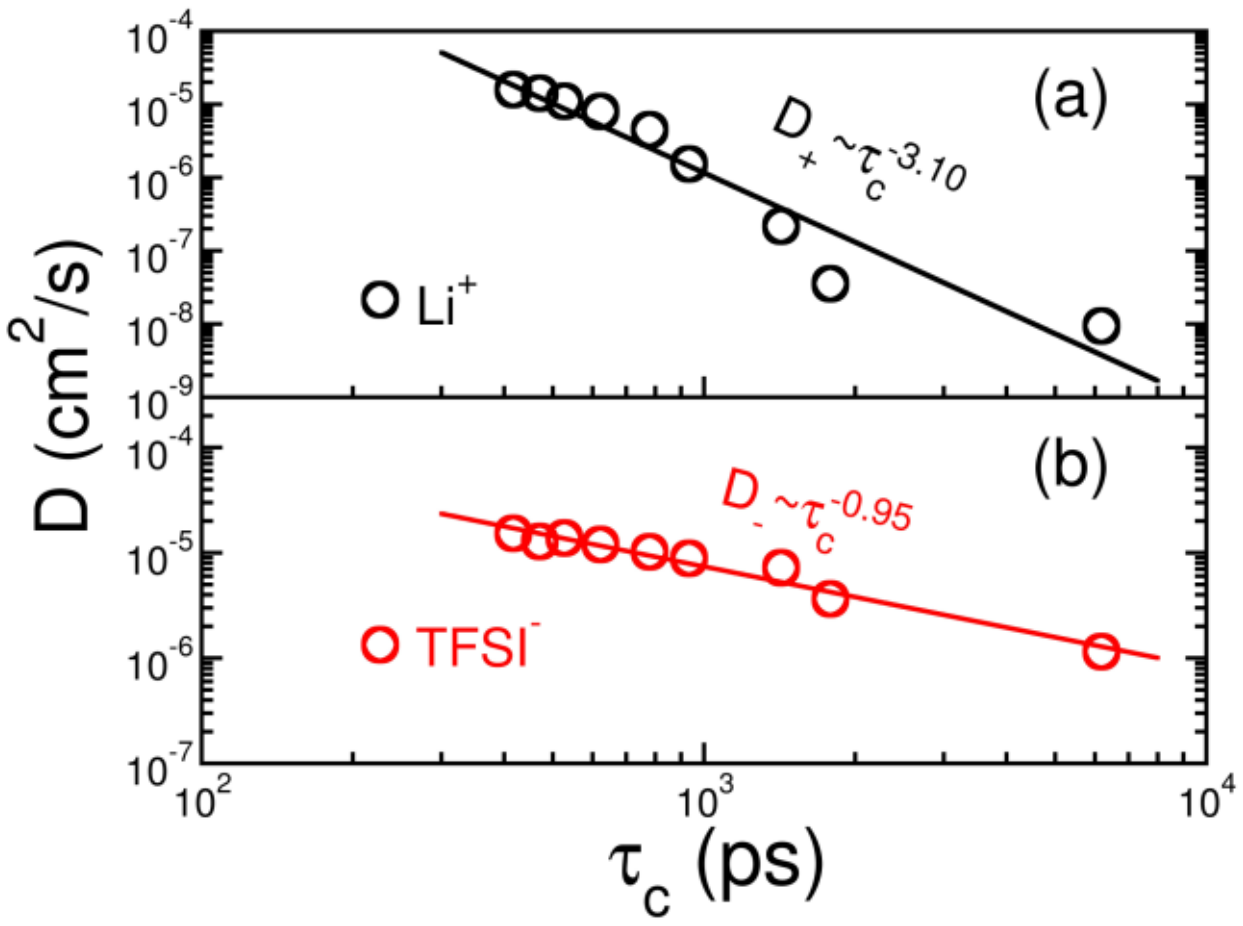

**Figure 6**: Correlations between (a) diffusion coefficient of $Li^+$ and ion-pair relaxation timescales, and (b) diffusion coefficient of $TFSI^-$ and ion-pair relaxation timescales.

The above analysis and results of **Figure 6** point to a surprising feature unique to the pectin-EC-LiTFSI electrolytes that both validity and breakdown of the $D\sim \tau_c^{-1}$ relation for a given type of ionic species. To the best of our knowledge, no soft matter or battery electrolyte material exhibits this unique feature, offering the flexibility of designing a new class of electrolytes for optimizing either cation or anion conductivity.

Now, the question is, what can we expect from this relative variation of diffusivity on ionic conductivity for different ions? As the ionic diffusivity of $Li^+$ ions is ~100 times less than that of the $TFSI^-$ ions, we can expect a negligible contribution towards ionic conductivity arising from $Li^+$ ions for higher wt. %s of pectin chains. Therefore, our designed electrolytes might behave as single-ion conductors. We will expand on this phenomenon in the following sections.

**Ionic Conductivity of Pectin-EC-LiTFSI Electrolytes**

The most crucial parameter in determining the suitability of an electrolyte for rechargeable batteries is its ability to conduct ionic charges efficiently. Ionic conductivity quantifies how well ions are transported in the designed electrolyte. Electrolytes offering the least resistance to the conduction of ions are highly desired for the design of rechargeable batteries. Designing a battery electrolyte with superior ionic conductivity and high viscosity simultaneously is the holy grail of solid polymer electrolyte research. The Nernst-Einstein (NE) equation, $\sigma_{NE} = \frac{e^2}{Vk_BT}(N_{Li}z_{Li}^2\overline{D}_{Li} + N_{TFSI}z_{TFSI}^2\overline{D}_{TFSI})$[82,83] for ionic conductivity can give excellent results for dilute solutions because of the highly uncorrelated motion of ions; in the NE equation, $e$ is the electronic charge, $V$ is the volume, $k_B$ is the Boltzmann constant, and $T$ is the absolute temperature. The symbols $N_\alpha$, $Z_\alpha$, $D_\alpha$, $\alpha = Li, TFSI$ indicate the number, ionic charge, and diffusion coefficient of $Li^+$ ions and $TFSI^-$ ions, respectively. However, the NE equation provides only an upper limit for the ionic conductivity at higher salt concentrations since the ion correlations start to dominate to lower the ionic conductivity. For instance, Borodin et al.[45] reported that about 20-30% of the total ionic conductivity was reduced because of the correlated motion of ions at a relatively high salt concentration of EC:Li ratios of 20:1 and 10:1. Similarly, in a different electrolyte system (PEO-LiTFSI) at an intermediate level of salt concentration measured in terms of EO:Li ratios of 39:1 and 20:1, about 5-10 % of the correlations were reported.[84]

Since the concentration of LiTFSI salt is considerably low (EC:Li=48:1) in all of our systems compared to the literature,[5,45] the cation-anion correlations will not be significant. Therefore, the Nernst-Einstein relationship can serve as a decent option for understanding the qualitative behavior of ionic conductivity of pectin-EC-LiTFSI electrolytes.[85]

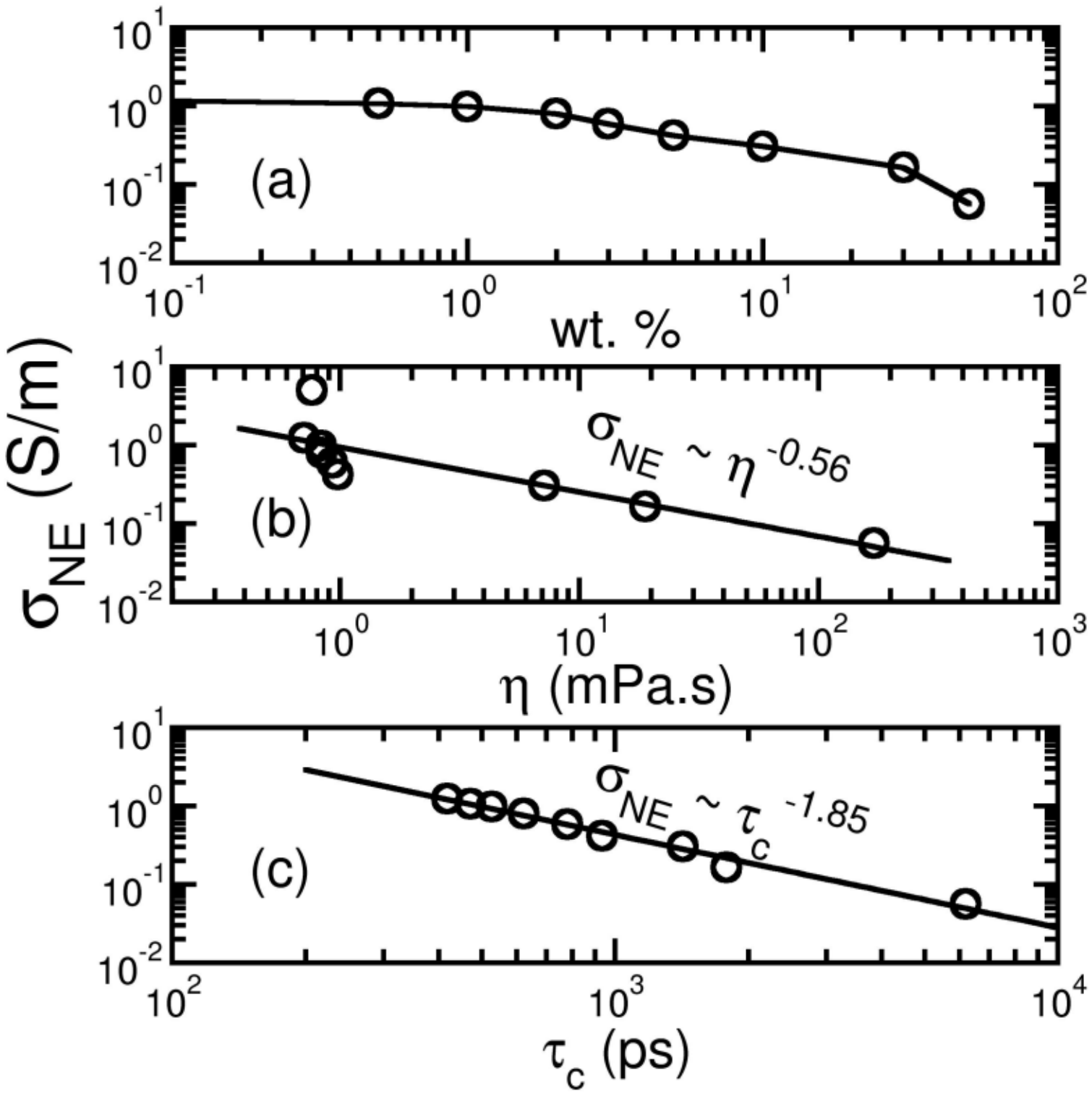

**Figure 7**: Nernst-Einstein ionic conductivity as a function of the (a) wt. % of pectin loading, (b) viscosity, and (c) ion-pair relaxation time.

As shown in **Figure 7**, the Nernst-Einstein conductivity of pure EC is 0.0123 S/cm, which agrees well with experimental ionic conductivity[86] and previous calculations based on MD simulations.[45] From our calculations, we observed a decrease in ionic conductivity similar to ionic diffusivities with the loading of pectin. The lowering of ionic conductivity with the addition of pectin is a direct consequence of the stronger interaction of lithium ions with the electronegative charge groups abundant in pectin chains. When the EC-LiTFSI system is loaded with 10 wt. % of pectin, the ionic conductivity decreases by a factor of four. Despite the conductivity being lower in higher wt. %s of pectin, we expect the ionic correlations to be insignificant, as clearly suggested by P(n) that the formation of smaller ionic clusters over the larger ones is more probable at high loadings of pectin. Therefore, true ionic conductivity will have fewer detrimental effects contributed by the correlated motion of ions. Further, since the rate of decrease of the diffusion coefficient of $Li^+$ ions is higher than that of $TFSI^-$ ions, the $TFSI^-$ ions contribute largely to the total ionic conductivity of the pectin-loaded EC-LITFSI systems. Our systems may be considered excellent ion conductors if we can increase the ionic conductivity of cations. Future research can focus on increasing the ion conductivity of cations by increasing the transference of cations through the electrolytes (see **Figure S9**).

Similar to the mechanisms examined for ionic diffusivities, we analyze the correlations between NE conductivity and viscosity/ion-pair relaxations and find that the viscosity and ion-pair relaxation timescales highly influence the Nernst-Einstein conductivity. Interestingly, we find significant differences between $\sigma_{NE}$ vs. $\eta^{\lambda}$ and $\sigma_{NE}$ vs. $\tau_c^{\lambda}$ as shown in **Figures 7 (b** and **c).** Explicitly, we observe that the NE conductivity depends on viscosity as $\sigma_{NE} \sim \eta^{-0.56}$ while on ion-pair relaxation timescales as $\sigma_{NE} \sim \tau_c^{-1.85}$, implying that (i) there exists distinct transport mechanisms for ionic conductivity, and (ii) that neither of the $\eta$ and $\tau_C$ correlates to NE conductivities. Despite the lower correlated motion of ions prevailing at a low salt concentration of EC:Li=48:1, the above results suggest that the respective ion-pair relaxation times influence ionic conductivity and that the existence of smaller ionic clusters can still negatively impact the ionic conductivity.

## Conclusions

In summary, we simulated a new class of solid polymer electrolytes derived from the biopolymer pectin chains that have enormous potential to convert the traditional liquid EC-LiTFSI electrolytes into solid polymer electrolytes and are environmentally compatible. Pectin is a natural biopolymer abundant in fruit extracts and agricultural products, exhibiting extraordinary properties such as high ionic sensitivity, biodegradability and non-toxicity due to its structural complexity. The anionic polysaccharide units consist of hydroxyl-rich molecular groups and can interact strongly with the cations in the electrolyte system. Due to these unique properties, we envisage pectin as a potential candidate to serve as a polymeric host for lithium ions and help them transport in the electrolyte material.

Our simulations predict that pectin has excellent ion solvating capabilities, as good as EC and polyethylene oxide chains, because of its structural uniqueness. We studied the structural properties by calculating the radial distribution function, coordination number, and different types of association probabilities. We found that the number of free lithium ions (i.e., lithium ions that are free from their coordination with $TFSI^-$ counterions) within the first coordination shell increases with the addition of pectin chains because of the increased coordination of lithium ions with the anionic groups of the pectin polymer chain. Further, the probability of association of pectin polymer polysaccharide units around lithium ions shows an increasing trend with the loading of pectin chains. Together, our simulations reveal that the pectin chains reduce the coordination numbers of $Li^+$ ions around the counterions (and vice-versa) because of stronger lithium-pectin interactions compared to lithium-TFSI interactions. As a result, the pectin promotes smaller ionic aggregates over larger ionic aggregates, contrary to those typically reported in liquid and polymer electrolytes. Formation of the smaller ionic aggregates over larger ionic aggregates is a promising feature of pectin and an interesting outcome for future battery electrolyte designs favoring the uncorrelated motion of ionic species and, thereby, efficient ionic conductivity.

The diffusion coefficient of ionic species was calculated using the respective MSDs based on 300 ns long trajectories at different wt. %s of pectin. The overall ion diffusivities are observed to follow a decreasing trend with the loading of pectin chains due to the strong coordination between $Li^+$ ions and polymer chains. However, the diffusion of both the ions at low pectin loadings (up to 3 wt. %) is almost unaffected by pectin and decreases by a factor of 100 for lithium-ion at higher loadings. Similarly, the Nernst-Einstein conductivity of pectin-EC-LiTFSI electrolytes monotonically decreases but slowly for pectin loadings up to 3 wt. %. For higher wt. % of pectin loadings, the ionic conductivity decreases rapidly due to the lower diffusion of $Li^+$ ions. As expected, from the transference number calculations, we see that the ionic conductivity depends heavily on the $TFSI^-$ ions, and the contribution of $Li^+$ ions towards conductivity is insignificant.

By comparing the diffusivity of ions and ion-pair relaxation timescales, we found interesting trends for the relation $D \sim \tau_c^{-\lambda}$. Explicitly, while the diffusivities of $TFSI^-$ ions obey the $D_- \sim \tau_c^{-1}$ relation with ion-pair relaxation timescales (i.e., $TFSI^-$ ions follow $D_- \sim \tau_c^{-0.95}$), the diffusivity of $Li^+$ ions do not obey the $D_+ \sim \tau_c^{-1}$ relation (i.e., $Li^+$ ions follow $D_+ \sim \tau_c^{-3.1}$). The excellent correlations found between diffusivities of $TFSI^-$ ions and ion-pair relaxation timescales are consistent with previous reports.[44,68,70,80] The lithium ion diffusivity does not correlate with ion-pair relaxation time because the polymeric pectin units trap the lithium ions. The Nernst-Einstein conductivity scales with ion-pair relaxation timescales as $\sigma_{NE} \sim \tau_c^{-1.85}$, revealing distinct transport mechanisms for ionic conductivity.

Our simulations established a surprising feature unique to the pectin-EC-LiTFSI electrolytes: Both the validity and the violation of the $D \sim \tau_c^{-1}$ relation for a given type of ionic species for the dependency of diffusion coefficient with $\tau_C$. To the best of our knowledge, no soft matter or battery electrolyte material exhibits such a unique feature, offering the flexibility of designing a new class of electrolytes for optimizing cation or anion conductivity. The reduced ion clustering, higher mechanical strength, and biodegradability are the futuristic advantages of biopolymer pectin that make it a promising alternative material to the state-of-the-art existing electrolyte technologies.

## Notes

The authors declare no competing financial interest.

## Acknowledgments

We thank Ananya Debnath for insightful discussions and critical feedback on the manuscript. The authors acknowledge the Computer Center of IIT Jodhpur, the HPC center at the Department of Physics, Freie Universität Berlin (10.17169/refubium-26754), for providing computing resources that have contributed to the research results reported in this paper. RRN acknowledges the support by Deutsche Forschungsgemeinschaft, Grant No. CRC 1349, Code No. 387284271, Project No. C04. SM acknowledges support from the SERB International Research Experience Fellowship SIR/2022/000786 and SERB CRG/2019/000106 provided by the Science and Engineering Research Board, Department of Science and Technology, India.

## Supporting information

The supplementary document contains **(S1)** The Visual Impression of the Simulated Systems (**S1.1**) The force field parameters for pectin polymer (Table **ST1:** bonded, non-bonded force field parameters for pectin polymer) (Table **ST2:** System details) **(S2)** Radial distribution function and Coordination number **(S3)** Ion association probability **(S4)** Mean squared displacement (Table **ST3:** Diffusion Coefficient Calculations from MSD curves), **(S5**) Calculation of Viscosity and Notes on Numerical Issues, **(S6)** Diffusion coefficient scaling with η and $\tau_c$ **(S7)** The ion-pair correlation function, and (**S8)** Transference Number with wt. % of Pectin.